# Charge carrier mobility degradation in graphene sheet under induced strain

Raheel Shah[†] and Tariq M. Mohiuddin

*Department of Physics, College of Science, Sultan Qaboos University, Muscat, Oman*

**Abstract:** Impact of induced strain on charge carrier mobility is investigated for a monolayer graphene sheet. Mobility is computed within Born approximation by including impurity scattering, surface roughness effects and interaction with lattice phonons. Unlike its sSi counterpart, strained graphene shows a drop in mobility with increasing strain. Main reason for this effect is decrease in Fermi velocity due to induced distortions in the graphene honeycomb.

† Corresponding author: raheelsh@squ.edu.om

## Introduction

Since its discovery the most explored area in the study of the 2D graphene sheet is its electronic properties. However, recently, a whole new wealth of information about graphene has emerged following the reports on its mechanical and other non-electronic properties. Graphene is the strongest material ever measured, with a breaking strength of $\sim 40\, N/m$ and Young's modulus ~1.0 $TPa$ [1]. Its thermal conductivity has been reported at a record value of ~5000 $Wm^{-1}K^{-1}$ [2]. Moreover, graphene was shown capable of withstanding reversible strain of as high as 20% [1]. Additionally, strain is a way forward towards bandgap engineering which is essential to convert graphene into semiconductor [3-4]. Collectively, these characteristics increase graphene's potential as a material of choice for NEMS and sensory device applications. Within these emerging novel properties, strain is considered very crucial as it promises whole new prospects for studying electronic transport [5] This is an idea being researched extensively theoretically [3, 6-14]. Although, fabricating an electronic device to extract Hall or field effect mobility in graphene while applying strain is a massive technological challenge, with recent efforts such as in ref. [15] it does not seem long before it takes shape. In the mean time however, several new non-electronic characteristics of graphene have emerged when studied under controllable applied strain [15-22].

The 2D nature of graphene means that strain laterally propagates within a layer without diminishing. Coupling this with its high resistance to both elastic deformation and breakage [1, 17], and additionally its high electronic quality [5, 23] provides motivation to explore changes in its electronic properties when put under strain. A further encouragement to this fact comes from theoretical reports on opening of a band gap in graphene for strains higher than 20% [3]. The importance of such a result for prospects of future graphene applications adds to our interest in further understanding its behaviour under applied strain. In this study we involve the complete spectrum of the typical nearest neighbour hopping parameters in the tight binding description of the density of states for a graphene lattice. The former are a set of three parameters which are reduced to one due to symmetry considerations [24], which is not applicable to graphene under strain. In the latter case the unit cell lattice vectors are strain dependent and they carry this dependency into the nearest neighbour hopping parameter description [3]. This enables one to describe the density of states and ultimately the Fermi velocity as a function of strain [25]. Considering three scattering mechanisms namely: remote impurity, phonon and surface roughness, we determine the Fermi velocity in the close vicinity of the Dirac point, and ultimately the conductivity dependent mobility of the charge carriers. Our results show that for applied strain along the special crystallographic orientations of Zigzag (*z*) and Armchair (*a*) the mobility of charge carriers is particularly degraded in the interval $20\% \leq \eta \geq 10\%$, the latter being the limit of strain used in our simulations. The reduction is more prominent in the *z* direction to that in the *a*. Temperature dependent investigations revealed a significant drop in mobility midway of the above mentioned range of strain only in the *z* direction.

# 1. Electronic Properties of Graphene

Graphene is the first known stable 2D material [23]. It is an allotrope of carbon composed of periodically arranged hexagons in a 2D one-atom thick infinite sheet. It is also considered as a semimetal with zero bandgap. Figure 1 illustrates a section of the infinite hexagonal network. Some prominent associated parameters are also sketched.

In equilibrium, distance between two adjacent carbon atoms is $a_o = 1.42 \overset{o}{A}$. Since each atom is shared by three adjacent hexagons thus the unit cell encloses one-third of each atom this leads to two atoms per unit cell. These two atomic sites are denoted by $A$ (filled circles) and $B$ (empty circles). Each atom $A(B)$ has three nearest neighbours $B(A)$ and six next nearest neighbours $A(B)$. The two primitive lattice vectors are $\mathbf{a} = a(1,0)$ and $\mathbf{b} = a(-1/2, \sqrt{3}/2)$, where $a = \sqrt{3} a_o$ while the lattice vectors joining site $A$ to site $B$ are denoted by $\boldsymbol{\delta}_i$ ($i = 1, 2, 3$).

Graphene's electro-magneto properties are sensitive to the edge effects [26-27], in particular along the two prominent directions viz. Zigzag ($z$) and Armchair ($\mathcal{A}$), depicted with broken lines in fig. 1. The coordinate axes can always be chosen such that x-axis is aligned along the $z$ orientation. Angle $\vartheta$ represents arbitrary vector directed in between $z$ and $\mathcal{A}$ orientations.

Each carbon atom in graphene possesses four valence electrons. The three in-plane $\sigma$ orbitals are tightly bound to neighbouring atoms. The fourth loosely bound $\pi$ orbital is perpendicular to the sheet and contributes to the electrical conductivity [28]. In the tight-binding model (TB) the energy bands in terms of nearest neighbour $\pi$ orbital hopping integrals ($t_i$, in fig. 1) is given by [13]:

$$E = \pm \left| t_2 + t_3 e^{-i\mathbf{k}\cdot(\mathbf{a}+\mathbf{b})} + t_1 e^{-i\mathbf{k}\cdot\mathbf{b}} \right| \quad (1.1)$$

where $\pm$ signs are for conduction and valence bands, respectively. $\mathbf{k} = (k_x, k_y)$ is the 2D wave-vector associated with charge carriers with energy $E$. First Brillouin zone (BZ) of the graphene unit cell is a hexagon itself.

The $\pi$ orbitals of the valence and conduction bands cross at two corners $K$ and $K'$ of the BZ also known as Dirac points. Charge carriers near the Dirac point behave like massless particles,

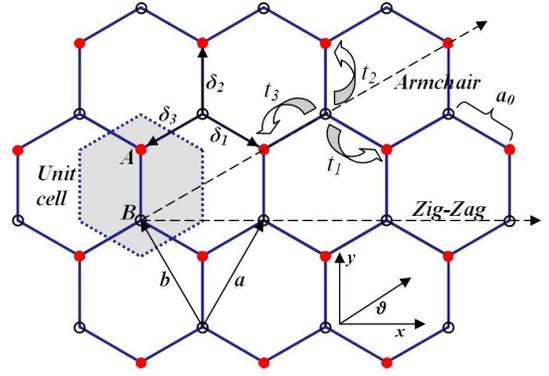

**Figure 1:** Part of infinite honeycomb network. Unit cell with two atoms $A$ and $B$ per cell. Hopping parameters $t_i$'s and bond lengths $\boldsymbol{\delta}_i$'s are also shown. Lattice primitive vectors are denoted by $\mathbf{a}$ and $\mathbf{b}$. Also shown are two distinctive directions in the network viz. zigzag and armchair. $a_o$ is the C-C distance. x-axis is aligned along zigzag orientation.

also for an intrinsic graphene Fermi energy is zero at these $K$ points.

Hamiltonian $H$ of $\pi$-bands near a Dirac point is described by the Dirac-Weyl equation [24, 29]:

$$H \Psi_{\pm\mathbf{k}} = \hbar v_F (\sigma_x k_x + \sigma_y k_y) \Psi_{\pm\mathbf{k}} = E \Psi_{\pm\mathbf{k}} \quad (1.2)$$

where $v_F$ is the Fermi velocity which is independent of charge carriers energy, $\sigma_x$ and $\sigma_y$ are Pauli spinors and $\hbar$ is the reduced Plank's constant. Eigenstates of Dirac-Weyl equation are given by plane wave as:

$$\Psi_{\pm\mathbf{k}} = \frac{1}{\sqrt{A}} e^{i(\mathbf{k}\cdot\mathbf{r})} \psi_{\pm\mathbf{k}} \quad (1.3)$$

where $A$ is the area of the system and $\mathbf{r} = (x, y)$ is the in-plane vector with

$$\psi_{\pm\mathbf{k}} = \frac{1}{\sqrt{2}} \begin{pmatrix} e^{i\theta_\mathbf{k}} \\ \pm 1 \end{pmatrix}; \theta_\mathbf{k} = \tan^{-1}\left(\frac{k_y}{k_x}\right) \quad (1.4)$$

The overlap of wave-function between initial and final states can easily shown to be [30]:

$$\left( \psi_{\pm\mathbf{k}}^* \cdot \psi_{\pm\mathbf{k}'} \right) = \frac{1}{2}\left( e^{i\theta} + 1 \right) \quad (1.5)$$

where $\theta$ is the angle between initial wave-vector $\mathbf{k}$ and the final scattered wave-vector $\mathbf{k}'$.

In equilibrium conditions the dispersion relation $E(\mathbf{k})$ near the Dirac point is given by utilizing equation (1.1) and (1.2) as [28]:

$$E = \pm \hbar v_F |\mathbf{k}| \quad (1.6)$$

Note at equilibrium $t_1=t_2=t_3 = 3.03\ eV$ [30] and $\delta_1 = \delta_2 = \delta_3 = a_o$

Electric transport is affected by the number of available vacant states in the system. In case of graphene, density of states (DoS) defined as $D(E) = \frac{1}{A}\sum_{\mathbf{k}'} \delta[E(\mathbf{k}) - E(\mathbf{k}')]$ can easily shown to be:

$$D(E) = \frac{g_v g_s}{2\pi} \frac{E}{(\hbar v_F)^2} \quad (1.7)$$

where $g_v = 2$ and $g_s = 2$ are introduced as the valley ($K \& K'$ points) and spin degeneracy, respectively. In terms of hopping parameter $t_o$ and C-C distance $a_o$, the Fermi velocity is given by $v_F = \frac{3 t_o a_o}{2\hbar}$, which turns out to be around 300 times less than the speed of light. Fermi velocity can alternatively be defined in terms of the unit cell area $A_c$ as [24]:

$$v_F = \frac{1}{\hbar}\sqrt{\frac{A_c \times \rho_t}{2}} \quad (1.8)$$

where $A_c = \frac{3\sqrt{3} a_o^2}{2}$ and $\rho_t = \sqrt{3} t_o^2$. This form is useful in the coming discussion.

## 1.1 Applied Strain

If strain is applied the lattice vectors $\delta_i$'s are modified as:

$$\delta_i = (\mathbf{I}_{2\times 2} + \boldsymbol{\eta})\cdot \delta_i^o \quad (1.1.1)$$

where $\delta_i^o$'s corresponds to the relaxed lattice vectors within the graphene unit cell. $\mathbf{I}_{2\times 2}$ is identity matrix of order 2 and the strain matrix $\boldsymbol{\eta}$ is given by [3]:

$$\boldsymbol{\eta} = \eta \begin{bmatrix} \cos^2\vartheta - \nu\sin^2\vartheta & (1+\nu)\cos\vartheta\sin\vartheta \\ (1+\nu)\cos\vartheta\sin\vartheta & \sin^2\vartheta - \nu\cos^2\vartheta \end{bmatrix} \quad (1.1.2)$$

where $\eta$ is the applied strain and $\nu = 0.14$ is the Poisson's ratio for graphene [31]. The deformation of lattice vectors $\delta_i$'s modify the hopping parameters $t_i$'s accordingly [3]:

$$t_i = t_o\, e^{-3.37(|\delta_i|/a_o - 1)} \quad (1.1.3)$$

The two distinctive directions in a graphene layer viz. $z$ and $a$ corresponds to $\vartheta = 0$ and $\vartheta = \pi/6$ respectively, while $\vartheta = \pi/3$ is the periodicity of system in $\vartheta$. It is also reported for $\eta < 20\%$ no change in bandgap is observed [3]. Simulations in this work will be bound to this limit.

DoS of the system is also altered by the induced strain. In general DoS is given by [25]:

$$D(E) = \frac{g_v g_s}{\pi} \frac{E}{(A'_c \rho'_t)} \quad (1.1.4)$$

where $A'_c$ is the area of the deformed unit cell and:

$$\rho'_t = \sqrt{(t_1^2 + t_2^2 + t_3^2)^2 - 2(t_1^4 + t_2^4 + t_3^4)} \quad (1.1.5)$$

comparing equation (1.1.4) with (1.7) one may define the "effective" Fermi velocity as $v'_F = \frac{1}{\hbar}\sqrt{\frac{A'_c \times \rho'_t}{2}}$ which reduces to equation (1.8) in the condition $\eta = 0$. The parameter $\rho'_t$ decreases monotonically with increasing strain.

The effective Fermi velocity reduces with increasing strain. Our simulations reveal an initial linear drop in $v'_F$ with increasing $\eta$ (slope of $\sim -1.2\times 10^8\ cm/s$ for $\eta \leq 10\%$) consistent with the recent study [7] and then relatively sharp drop in $v'_F$ for $\eta > 10\%$. Thus it is expected that with increasing strain the dc conductivity and the corresponding mobility will be degraded.

## 2. Transport Models in the Vicinity of Dirac Point

DC conductivity and the corresponding mobility is calculated using the rigorous linearization of the Boltzmann transport equation, given by in the absence of external magnetic field as [32]:

$$v_\mathbf{k}\cdot \nabla_\mathbf{r} f_\mathbf{k}^c - \frac{e\mathbf{E}_c}{\hbar}\cdot \nabla_\mathbf{k} f_\mathbf{k}^c = -\left.\frac{\partial f_\mathbf{k}^c}{\partial t}\right|_{scatt} \quad (2.1)$$

where $v_\mathbf{k}$ is the charge carrier's velocity and $E_c$ is the applied electric field. The electric field dependent distribution function $f_\mathbf{k}^c$ is expanded around the equilibrium value $f_\mathbf{k}$ such that $f_\mathbf{k}^c = f_\mathbf{k} + g_\mathbf{k}$, where $g_\mathbf{k}$ is assumed to be weak perturbation to the $f_\mathbf{k}$. Further in the Relaxation Time Approximation (RTA) the term

$$-\left.\frac{\partial f_{\mathbf{k}}^c}{\partial t}\right|_{scatt}$$ is approximated by $\frac{g_{\mathbf{k}}}{\tau_{\mathbf{k}}}$, where $\tau_{\mathbf{k}}$ is the relaxation time. It can be readily shown that [33]

$$\frac{1}{\tau_{\mathbf{k}}} = \sum_{\mathbf{k}'} W(\mathbf{k},\mathbf{k}')(1-\cos\theta) \quad (2.2)$$

The quantum mechanical scattering probability $W(\mathbf{k},\mathbf{k}')$ can be computed within Born approximation as:

$$W(\mathbf{k},\mathbf{k}') = \frac{2\pi}{\hbar}\left|V_{\mathbf{k}',\mathbf{k}}^n\right|^2 \delta(E_{\mathbf{k}'}-E_{\mathbf{k}} \mp \Delta E) \quad (2.3)$$

The Dirac delta function ensures the conservation of energy, $\Delta E$ is the change in energy, if any, while the matrix element $\left|V_{\mathbf{k}',\mathbf{k}}^n\right|$ is defined as:

$$\left|V_{\mathbf{k}',\mathbf{k}}^n\right| = \frac{1}{A}\int \psi_{\mathbf{k}'}^*(\mathbf{r}) V_s^n(\mathbf{r}) \psi_{\mathbf{k}}(\mathbf{r}) d\mathbf{r} \quad (2.4)$$

where $\psi_{\mathbf{k}}(\mathbf{r})$ is the wave-function associated with charge carrier and $V_s^n(\mathbf{r})$ is the perturbation potential of the n$^{th}$ type responsible for scattering in the system of area $A$

The electric current $\mathbf{J}$ in terms of carrier velocity and the distribution function $g_{\mathbf{k}}$ is given by [34]:

$$\mathbf{J} = \frac{g_v g_s}{A}\sum_{\mathbf{k}} e v_{\mathbf{k}} g_{\mathbf{k}} \quad (2.5)$$

with $g_{\mathbf{k}} = e\tau_{\mathbf{k}} v_{\mathbf{k}} \cdot \mathbf{E}_c \delta(\zeta - v_F \hbar k)$ is the solution of the linearized Boltzmann equation. $\zeta$ is the chemical potential in the system. Additionally, the effective scattering rate:

$$\frac{1}{\tau_{eff}(E)} = \sum_n \frac{1}{\tau^{(n)}(E)} \quad (2.6)$$

is utilized to compute electric conductivity $\sigma_e$ within Kubo-Greenwood transport formalism as:

$$\sigma_e = \frac{e^2 v_F^2}{2}\frac{\int E D(E) \tau_{eff}(E)(-\partial f/\partial E)dE}{\int E(-\partial f/\partial E)dE} \quad (2.7)$$

And finally the carrier mobility $\mu$ in terms of conductivity and carrier density $n_s$ is calculated as:

$$\mu = \frac{\sigma_e}{e n_s} \quad (2.8)$$

### 2.1 Remote Impurity Interaction

Charge impurities present in the substrate is one of the significant sources of mobility degradation in graphene [30, 35-36]. In $\mathbf{k}$-space the Coulomb scattering potential $V_s^C(\mathbf{r})$ is given by:

$$V_s^C(q) = \frac{2\pi e^2 e^{-dq}}{\kappa q} \quad (2.1.1)$$

where charged impurities in the substrate are assumed to be at a distance $d$ away from the graphene sheet, the scattered wave-vector $q = |\mathbf{k}-\mathbf{k}'| = 2k\sin\theta/2$ and $\kappa$ is the effective dielectric constant of the system. Inclusion of screening effect is vital in observing the impact of charge impurity. In particular for graphene the screened Coulomb potential results in the matrix element as:

$$\left|V_{\mathbf{k}',\mathbf{k}}^C\right| = \frac{V_s^C(q)}{\varepsilon(q)}\left(\psi_{\mathbf{k}'}^* \cdot \psi_{\mathbf{k}}\right) \quad (2.1.2)$$

where $\varepsilon(q)$ is the static dielectric function for a 2D graphene sheet. Starting from the Lindhard function and under random phase approximation (RPA) [37] $\varepsilon(q)$ is given in terms of polarization function $\Pi(q)$ as [30]:

$$\varepsilon(q) = 1 + \frac{2\pi e^2}{\kappa q}\Pi(q) \quad (2.1.3)$$

$$\Pi(q) = \frac{g_v g_s}{2\pi(v_F \hbar)}\int_0^\infty f_{\mathbf{k}}^+ dk - \int_0^{q/2} f_{\mathbf{k}}^+ \sqrt{1-\left(\frac{q}{2k}\right)^2} dk \quad (2.1.4)$$

where $f_{\mathbf{k}}^+ = f(E_{\mathbf{k}}) + f(E_{\mathbf{k}}+2\zeta)$ with $f(E_{\mathbf{k}})$ is assumed to be the Fermi-Dirac function.

Finally the scattering rate for the Coulomb interaction is the presence of $n_i$ charge centres per area is then derived as:

$$\frac{1}{\tau^{Col}} = \frac{2\pi}{\hbar}\left(\frac{n_i D(E)}{2\pi}\right) \\ \times \int_0^{2\pi}\left(\frac{1-\cos^2\theta}{2}\right)\left(\frac{V_s^C(q)}{\varepsilon(q)}\right)^2 d\theta \quad (2.1.5)$$

### 2.2 Surface Roughness Interaction

As with any surface and/or interface graphene-deposited on the substrate has a non-smooth surface. Impact of surface roughness on charge carrier mobility is extensively studied for quasi-2D structures [38-41] to name but a few. In this work the impact due to difference in the dielectric values at the graphene/substrate interface is taken into account. The interface randomness is, as usual, modelled by the autocovariance function between $\Delta$ -the rms height of the random interface "steps" and $\Lambda$ - the average width of the same fluctuation. For an exponential autocorrelation form the power spectrum density $|S(q)|^2$ is given by [40]:

$$|S(q)|^2 = \pi \Lambda^2 \Delta^2 (1 + q^2 \Lambda^2 / 2)^{-\frac{3}{2}} \quad (2.2.1)$$

In the presence of two different dielectric materials at the interface polarization charges are created. The potential induced by such polarization is given by [41]:

$$V_s^{pol}(q) = e \tilde{\varepsilon} E_{eff} e^{-qz_0} \quad (2.2.2)$$

Where the parameter $\tilde{\varepsilon} = (\varepsilon_g - \varepsilon_{sub})/(\varepsilon_g + \varepsilon_{sub})$ with $\varepsilon_g$ and $\varepsilon_{sub}$ denote the dielectric constants of the graphene and the substrate, respectively. Effective electric field $E_{eff}$ at the graphene side of the interface is defined here as $E_{eff} = \frac{e}{\varepsilon_g}(n_i + n_s)$, while $z_0$ is taken as $\Delta$ of the interface.

The mismatch of the dielectric constants at the interface also introduces image charges [39]. For sufficiently thick substrate the scattering potential associated with image charges is given by [41]:

$$V_s^{img}(q) = \frac{e^2 \tilde{\varepsilon} q^2}{16 \pi \varepsilon_g} \left( \frac{K_1(qz_0)}{qz_0} - \frac{\tilde{\varepsilon}}{2} K_0(qz_0) \right) \quad (2.2.3)$$

where $K_0$ and $K_1$ are modified Bessel functions of the second kind of order zero and one, respectively. Net impact of these two scattering sources is thus given by:

$$V_s^{SR}(q) = V_s^{pol}(q) + V_s^{img}(q) \quad (2.2.4)$$

Note that the net scattering strength not only depends on the dielectric properties of the substrate but more importantly on the difference between the dielectric properties of the two materials at the interface. Surface roughness induced scattering potential thus yields the scattering rate as:

$$\frac{1}{\tau^{SR}} = \frac{2\pi}{\hbar} \left( \frac{D(E)}{2\pi} \right) \times \int_0^{2\pi} \left( \frac{1 - \cos^2 \theta}{2} \right) \left( \frac{V_s^{SR}(q)}{\varepsilon(q)} \right)^2 |S(q)|^2 d\theta \quad (2.2.5)$$

### 2.3 Phonon Interaction

For ambient and higher temperature regime carrier-phonon interaction is a major cause of mobility degradation in electronic devices [42]. Low energy acoustic phonons are treated under elastic scattering approximation. It is reported that group symmetry forbids TA phonon modes to exist for graphene [36, 43], therefore only LA mode of the spectrum is taken into account with relatively stronger coupling (see discussion below).

The scattering potential associated with acoustic phonons is given by [33, 44]:

$$V_s^{ac}(q) = \frac{D_{ac}}{v_{ph}} \sqrt{\frac{k_B T}{2 \rho_g A}} e^{i(\mathbf{q} \cdot \mathbf{r} - \omega t)} \quad (2.3.1)$$

where $D_{ac}$ is the deformation potential of the graphene lattice, $\rho_g$ is the surface density of the system. Phonon velocity is denoted by $v_{ph}$, $k_B$ is Boltzmann constant and $T$ is the temperature in Kelvin's scale. Here linear phonon dispersion is assumed (i.e. acoustic phonon frequency $\omega_q^{ac} = v_{ph} q$) and equipartition approximation is applied which is valid for moderate to higher temperature regimes.

Impact of screening on electron-phonon interaction is long debated however via simulations the ineffectiveness of screening in case of electron-phonon scattering is concluded [45], following this, dynamical screening is not included here.

Finally the acoustic phonon scattering rate is derived through the evaluation of matrix element as:

$$\frac{1}{\tau^{ac}} = \frac{2\pi}{\hbar} \left( \frac{D_{ac}^2 k_B T}{8 v_{ph}^2 \rho_g} \right) D(E) \quad (2.3.2)$$

Non-polar optical phonons are treated inelastically in the simulations. Similar to acoustic phonon spectrum only LO phonons contribute towards scattering mechanism. The interaction potential in this case is given by [33]:

$$V_s^{op}(q) = D_{op}\sqrt{\frac{\hbar}{2\rho_g A\omega_q^{op}}}$$
$$\times \sqrt{\left(N_q + \frac{1}{2} \pm \frac{i}{2}\right)}e^{i(\mathbf{q}\cdot\mathbf{r}-\omega t)} \quad (2.3.3)$$

where $N_q$ is the Bose-Einstein distribution function, $D_{op}$ is the deformation field associated with vibrating lattice sites, $i = -1$ is for phonon absorption (upper sign) and $i = +1$ is for emission process (lower sign). Optical phonon frequency is denoted by $\omega_q^{op}$ and in the dispersion region of interest it bears a constant value $\omega_o$, independent of the transferred wave-vector $q$.

The interaction potential leads to the scattering rate as:

$$\frac{1}{\tau^{op}} = \frac{2\pi}{\hbar}\left(\frac{\hbar D_{op}^2}{4\rho_g \omega_o}\right)\left(N_o + \frac{1}{2} \pm \frac{i}{2}\right) \quad (2.3.4)$$
$$\times D(E \pm \hbar\omega_o)\Theta(E \pm \hbar\omega_o)$$

where the Heaviside step function $\Theta(x)$ is introduced to account for only physically possible scattering events.

## 3. Results

For the Coulomb interaction the location of charge centres in the substrate away from the interface is taken as $d \approx 1\,nm$ [35] while the charge density $n_i$ is assumed to be around $1.5 \times 10^{11}\,\#/cm^2$.

Unless graphene surface topology is thoroughly investigated the autocorrelation of the step width $\Lambda$ and rms step height $\Delta$ will be used as fine tuning parameters to simulate observed mobility. With the exponential autocovariance model used here, $\Lambda$ and $\Delta$ parameters are taken as $1.0\,nm$ and $0.5\,nm$, respectively. Relative dielectric constant of graphene is assumed to be 5.7 [36].

In the phonon interaction model the value of acoustic deformation potential $D_{ac}$ is not settled yet. In literature values as small as $4.75\,eV$ to as large as $30\,eV$ are quoted [36, 46] (and references within). Ignoring the anisotropy of the deformation potential a single constant value $D_{ac} \sim 20\,eV$ is used in this work. Phonon velocity $v_{ph}$ of the LA branch is set to $2.0 \times 10^6\,cm/s$ and graphene density is taken as $\rho_g = 7.6 \times 10^{-8}\,g/cm^2$ [47]. The deformation field constant $D_o$ appearing in optical phonon model is assumed to have the strength of $2.0 \times 10^9\,eV/cm$ while LO phonon is associated with energy of $\hbar\omega_o = 152\,meV$ [36].

Figure 2 shows, in the absence of any strain, the relative scattering strength of the three scattering mechanisms studied here. Both phonon and surface roughness scattering rates tend to increase with carrier's energy.

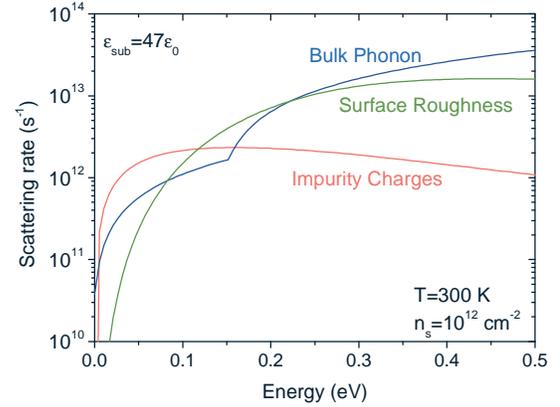

**Figure 2:** Scattering rate for lattice phonon, impurity charges and surface roughness interaction at room temperature. For electron energy > 0.15eV optical phonon emission process is possible and thus a sharp increase in phonon scattering rate is observed. Coulomb scattering rate initially increases with energy and then drops for sufficiently high carrier's energy.

In order to benchmark the transport models and their respective parameters, simulations are performed and compared with the reported measured/extracted mobilities in ref. [48].

Figure 3 shows the comparison of simulated and the measured data. Simulations are performed for substrate with dielectric constant of $47\varepsilon_0$ at room temperature. As it can be seen the modelled mobility reasonably follows the reported trend and its magnitude.

Inclusion of high dielectric ($\varepsilon_{sub}$) constant possibly have diverse effects on carrier mobility. Firstly, strength of impurity charge centers is certainly reduced due to screening and thus mobility is expected to increase. On the other hand possible soft optical phonon modes in the substrate and relatively higher amount of charge impurities present in the dielectric could result in low mobilities.

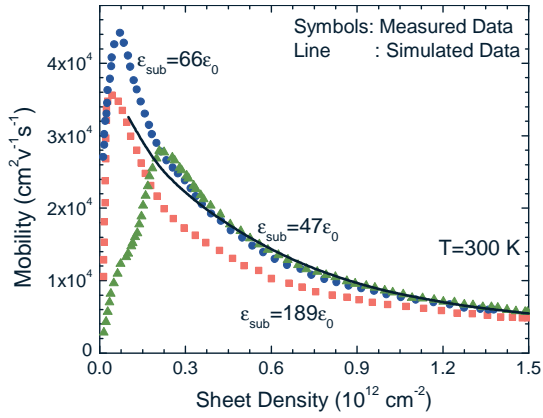

**Figure 3:** Simulations are benchmarked against the measured mobilities as a function of sheet density. Experimental data are from ref. [48].

Interface roughness is not necessarily same for each selected substrate, in addition, theoretically the difference between $\varepsilon_{sub}$ and $\varepsilon_g$ alters the impact of surface roughness provided $\Lambda$ and $\Delta$ parameters are assumed to be same for different dielectrics. Under identical conditions our simulations reveal higher mobilities for graphene on Silicon oxide ($\varepsilon_{sub} = 3.9\varepsilon_0$) as compared to Dimethyl sulfoxide ($\varepsilon_{sub} = 47\varepsilon_0$), this is due to the reduced scattering potential $V_s^{SR}(q)$ associated with SR.

Next, simulations are performed with strain induced in the graphene sample. Figure 4 depicts the results obtained for three sets of induced strain, both along $z$ and $\mathcal{A}$ orientations. For strain induced around 10% the decrease in mobility along $z$ and $\mathcal{A}$ orientations is almost same but as $\eta$ is increased to around 20% impact of strain is more prominent along $z$ orientation.

Figure 5 gives the mobility profiles as a function of induced strain for low, moderate and elevated temperature range. Base substrate is $SiO_2$ in this case. At low temperature (77 K) contribution of phonons in scattering mechanism is negligible and only impurity charges and surface roughness play their dominant parts. In this low temperature regime mobility very nominally increases as the induced strain in increased below 13%. This is due to the reason that the net scattering rate stays almost constant for low $T$ and $\eta$ values but the Fermi derivative term $(-\partial f / \partial E)$, appearing in conductivity expression, increases relatively rapidly and thus effective conductivity shows a positive slope in this region.

## 4. Conclusion

In this study charge carrier mobility in a monolayer graphene sheet is computed under induced strain both along $z$ and $\mathcal{A}$ edges. It is predicted that mobility and hence dc conductivity will degrade with increasing strain.

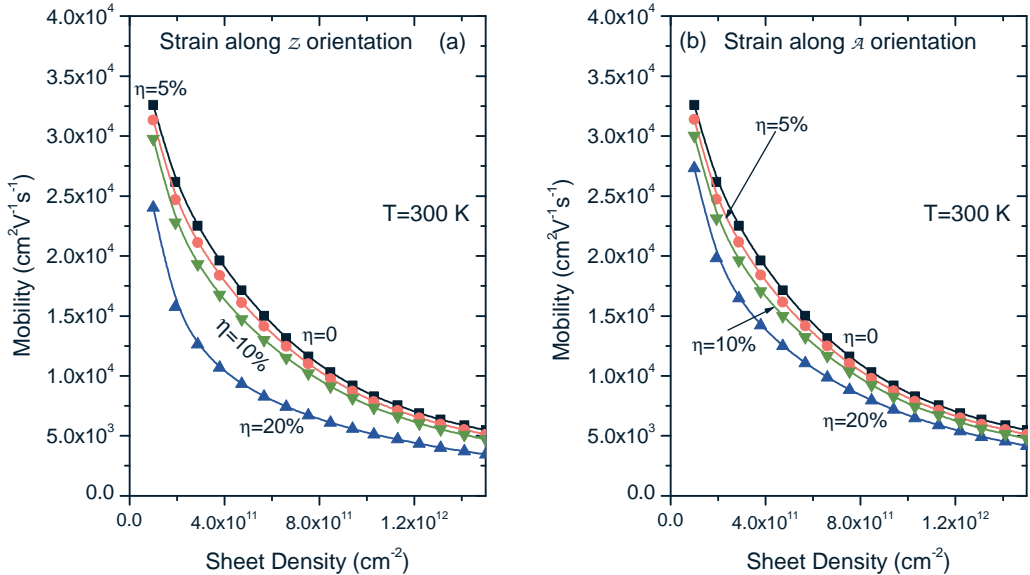

**Figure 4:** Impact of strain on simulated mobilities against increasing gate voltage. Mobility degradation is prominent for $20\% \leq \eta \geq 10\%$. (a) $z$ direction, (b) $\mathcal{A}$ direction. The base substrate is Dimethyl Sulfoxide ($\varepsilon_{sub} = 47\varepsilon_0$).

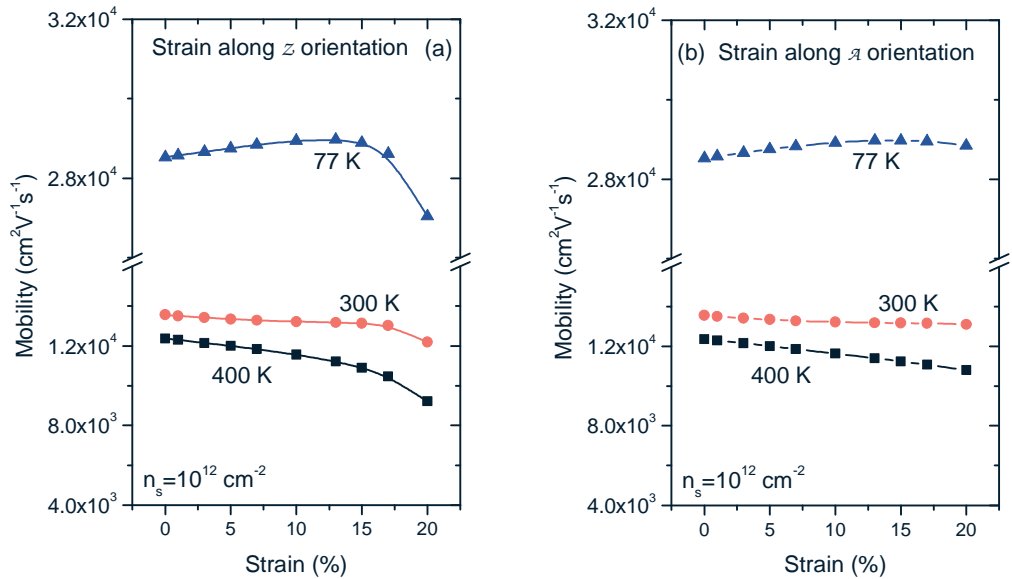

**Figure 5:** strain dependent simulated mobilities for low to high temperature regimes. Base substrate is SiO$_2$ and sheet density is $10^{12}\ cm^{-2}$.

Prime reason for this observation is the decrease in the Fermi velocity which in turn is inversely proportional to the available DoS in the graphene system. Fermi velocity is computed in terms of hopping parameters which are functions of distorted bond lengths between C-C atoms. Three main scattering mechanisms are included in the simulations viz. remote impurity, interface roughness and lattice phonon (both acoustic and optical) interaction.

## Acknowledgments

We would like to thank Prof. Ram Singh for useful discussions and guidance throughout this work. Authors also acknowledge the facilities provided by Sultan Qaboos University for this study.